\documentclass{IEEEtran}

\usepackage[english]{babel}
\usepackage{cite}
\usepackage{float}
\usepackage{caption}
\usepackage{type1cm} % type1 computer modern font
\usepackage{graphicx} % advanced figures
\usepackage{xspace} % fix space in macros
\usepackage{balance} % to better equalize the last page
\usepackage{booktabs} % nicer tables
\usepackage{multirow} % multi rows for tables
\usepackage[font={bf}, tableposition=top]{caption} % captions on top for tables
\usepackage{subcaption} % subfloats
\usepackage{bold-extra} % bold + {small capital, italic}
\usepackage{algorithm}
\usepackage{algorithmic}
\usepackage{microtype} % compress text
\usepackage{siunitx} % \num for decimal grouping
\usepackage{xfrac} % nicer slanted fractions
\usepackage{amsmath}
\usepackage{amssymb}
\usepackage{mathtools}
\usepackage{amsfonts}
\usepackage{amsthm}
\usepackage{bbm}
\usepackage[hyphens]{url} % handle long urls
\usepackage[bookmarks, pdftex, colorlinks=true]{hyperref} % clickable references
\usepackage{csquotes}
\usepackage[show]{chato-notes} % notes
\usepackage{bm}
\usepackage{xspace}
\usepackage{physics}
\usepackage{newunicodechar}
\newunicodechar{⟓}{\ensuremath{\uplus}}
\usepackage{tikz}
\usetikzlibrary{positioning}
\usetikzlibrary{bayesnet}
\usetikzlibrary{arrows}
\usetikzlibrary{shapes}
\usetikzlibrary{fit}
\usepackage{tikz-cd}
\usepackage{bbold}
\usetikzlibrary{calc, positioning, arrows.meta, shapes.geometric, fit, backgrounds}
\usepackage{pgf}
\usepackage{tikz-3dplot}
\usepackage{tcolorbox}

\usepackage[dvipsnames]{xcolor}
\hypersetup{
   bookmarks, pdftex,
   colorlinks=true,
   pagebackref=true, backref=page,
   linkcolor={red!50!black},
   filecolor={green!50!black},
   citecolor={green!50!black}, 
   urlcolor={blue!80!black},
}

\graphicspath{{fig}}

\theoremstyle{plain}
\newtheorem{theorem}{Theorem}
\newtheorem{proposition}[theorem]{Proposition}

\newcommand{\F}{\ensuremath{\mathbf{F}}\xspace}

\usepackage{stackengine}

\definecolor{mypurple}{RGB}{254, 68, 218}

\title{Learning Dirac Spectral Transforms \\for Topological Signals}
\author{
\IEEEauthorblockN{Leonardo Di Nino$^{1,3}$, Tiziana Cattai$^1$, Sergio Barbarossa$^1$, Ginestra Bianconi$^2$ and Paolo Di Lorenzo$^{1,3}$ \\ \smallskip}
\IEEEauthorblockA{
$^1$Dept. of Information Engineering, Electronics, and Telecommunications, Sapienza University of Rome, Italy \\
$^2$School of Mathematical Sciences, Queen Mary University of London, UK.\\
$^3$ Consorzio Nazionale Interuniversitario per le Telecomunicazioni, Parma, Italy \smallskip\\
Email: \{leonardo.dinino, tiziana.cattai, sergio.barbarossa, paolo.dilorenzo\}@uniroma1.it}, g.bianconi@qmul.ac.uk
\thanks{This work was supported by the European Union under the Italian National Recovery and Resilience Plan (NRRP) of NextGenerationEU, partnership on “Telecommunications of the Future” (PE00000001 - program “RESTART”), and by the Horizon SNS JU project 6G-GOALS under grant no. 101139232.}\vspace{-.4cm}}

\begin{document}
\maketitle

\begin{abstract}
The Dirac operator provides a unified framework for processing signals defined over different order topological domains, such as node and edge signals. Its eigenmodes define a spectral representation that inherently captures cross-domain interactions, in contrast to conventional Hodge–Laplacian eigenmodes that operate within a single topological dimension.
In this paper, we compare the two alternatives in terms of the distortion/sparsity trade-off and we show how an overcomplete basis built by concatenating the two dictionaries can provide better performance with respect to each approach. Then, we propose a parameterized nonredundant transform whose eigenmodes incorporate a mode-specific mass parameter that captures the interplay between node and edge modes. Interestingly, we show that learning the mass parameters from data makes the proposed transform able to achieve the best distortion-sparsity trade-off with respect to both complete and overcomplete bases. 

\end{abstract}
\begin{keywords}
Graph Signal Processing, Dirac operator of networks, data-driven transform learning, topological signals. \vspace{-.2cm}
\end{keywords}

\section{Introduction}
Over the years, significant interest has grown in developing algorithms of algorithms for processing and learning from signals defined over the nodes of a network.
The whole machinery of Graph Signal Processing (GSP) \cite{sandryhaila2013discrete} revolves mainly around Laplacian operators \cite{smola2003kernels} as key enablers for spectral and convolutional methods based on the localization properties of the Graph Fourier Transform.
Also, when moving to higher order topological structures to process edge signals, Hodge-Laplacians are still the cornerstone for both linear \cite{Barbarossa_2020,isufi2025topological} and deep neural methods \cite{yang2022simplicial}.
However, Laplacian operators inherently treat signals within a single topological dimension at a time. 
In contrast, the Dirac operator of networks \cite{bianconi2021higher,bianconi2021topological} naturally couples multiple topological domains, enabling the joint processing of information across different orders. 
This property makes the Dirac operator particularly appealing for studying network dynamics \cite{calmon2023local,carletti2025global},
topological data analysis  \cite{baccini2022weighted,wee2023persistent} and neural architecture design \cite{mey2024dirac,battiloro2024generalized}.
% and modeling systems where different domains are intrinsically entangled by physical principles, such as molecular representation .

Algorithms and methods derived from the Dirac equation have demonstrated remarkable performance in processing tasks.
In particular, building on its constitutive properties, it enables the design of filtering methods for topological signals, generalizing Tikhonov-like approaches to signals localized around a principal Dirac eigenmode \cite{calmon2023dirac} or by recursively capturing subsequently the main contributions aligned to multiple modes \cite{wang2025dirac}.
However, to date, it remains unclear under which structural and signal-dependent conditions Laplacian or Dirac bases are more appropriate for processing graph and topological signals, including tasks such as representation, filtering, or sampling. While Laplacian operators provide dimension-wise smooth decompositions, the Dirac operator captures cross-dimensional differential couplings; however, real-world signals may only partially adhere to either model. As a result, the trade-off between spectral localization, sparsity, and topological entanglement in graph signals is not yet fully understood.

\noindent\textbf{Contributions.} In this work, we address the aforementioned gap by systematically investigating the conditions under which Dirac-based representations offer advantages over Laplacian ones. 
Specifically, the main contribution of this work is threefold: \emph{(i)} we provide a systematic study of spectral localization properties for joint node-edge signals on networks, clarifying the trade-offs between sparsity, smoothness, and cross-domain coupling; \emph{(ii)} we construct an overcomplete Laplacian–Dirac frame that allows for the joint analysis of topological signals, providing enhanced flexibility in representation and mode selection; \emph{(iii)} we introduce an adaptive spectral transform that generalizes both Laplacian- and Dirac-based signal processing, together with a data-driven framework to infer it from observations; the proposed transform learns a sparse mixture of Dirac operators with varying mass parameters, enabling adaptive coupling across topological domains. Numerical experiments on synthetic and real-world datasets confirm the effectiveness of the proposed framework, highlighting its competitiveness with respect to standard Dirac- and Laplacian-based approaches.
\vspace{-.1cm}
\section{Dirac--Laplacian Representation}\label{sec:2}

\subsection{Representation of Topological Signals}

We consider a graph $\mathcal{G}(\mathcal{V},\mathcal{E})$ with $|\mathcal{V}|=V$ nodes and $|\mathcal{E}|=E$ edges, and node/edge signals $\mathbf{x}_0\in\mathcal{C}^0\cong\mathbb{R}^V$, $\mathbf{x}_1\in\mathcal{C}^1\cong\mathbb{R}^E$. A \textit{topological spinor} $\mathbf{s}$ is the composite vector \cite{bianconi2021topological}:
\begin{equation}
\mathbf{s}=\begin{pmatrix} \mathbf{x}_0 \\ \mathbf{x}_1 \end{pmatrix}\in\mathcal{C}^0\oplus\mathcal{C}^1\cong\mathbb{R}^{V+E}.
\end{equation}
The two domains are related by the \textit{discrete gradient} $\delta:\mathcal{C}^0\rightarrow\mathcal{C}^1$ and \textit{divergence} $\delta^*:\mathcal{C}^1\rightarrow\mathcal{C}^0$ \cite{grady2010discrete}, acting element-wise as
\begin{equation}\label{eq:gradient}
    (\delta\mathbf{x}_0)_e=\mathbf{x}_{0}[v]-\mathbf{x}_0[u], \quad\forall e:u\trianglelefteq e\trianglerighteq v
\end{equation}
\begin{equation}\label{eq:divergence}
    (\delta^*\mathbf{x}_1)_v=-\!\!\sum_{e:v\trianglelefteq e}\!\!\mathbf{x}_1[e]+\!\!\sum_{e:v\trianglerighteq e}\!\!\mathbf{x}_1[e], \quad \forall v \in \mathcal{V},
\end{equation}
where $\trianglelefteq,\trianglerighteq$ denote oriented incidence relations. Both operators are encoded by the incidence matrix $\mathbf{B}$ ($\mathbf{B}[n,e]=1$ for $n\trianglerighteq e$, $-1$ for $n\trianglelefteq e$, $0$ otherwise), so that $\delta\mathbf{x}_0=\mathbf{B}^\top\mathbf{x}_0$ and $\delta^*\mathbf{x}_1=\mathbf{B}\mathbf{x}_1$. Their composition yields the graph Laplacian $\mathbf{L}_0\mathbf{x}_0=\mathbf{B}\mathbf{B}^\top\mathbf{x}_0$ (divergence of the gradient) and the $1$-Hodge Laplacian $\mathbf{L}_1\mathbf{x}_1=\mathbf{B}^\top\mathbf{B}\mathbf{x}_1$ (gradient of the divergence), whose spectra reveal topological structure and enable domain-aware processing \cite{chung1997spectral,ortega2022introduction}. However, each Laplacian acts on a single domain at a time, whereas the underlying constitutive relation suggests joint representation across contiguous domains.

To this end, the Dirac operator of networks couples the differential relations \eqref{eq:gradient}--\eqref{eq:divergence} \cite{bianconi2021topological,calmon2023dirac,wang2025dirac}. The operator $\mathbf{D}:\mathcal{C}^0\oplus\mathcal{C}^1\rightarrow\mathcal{C}^0\oplus\mathcal{C}^1$ maps one domain into the other, and its square recovers the super Laplacian $\mathbf{L}_{\mathcal{G}}=\mathrm{blkdiag}(\mathbf{L}_0,\mathbf{L}_1)$:
\begin{equation}
    \mathbf{D}\mathbf{s}=\begin{pmatrix} \mathbf{0} & \mathbf{B} \\ \mathbf{B}^\top & \mathbf{0} \end{pmatrix}\!\mathbf{s} = \begin{pmatrix} \mathbf{B}\mathbf{x}_1 \\ \mathbf{B}^\top\mathbf{x}_0 \end{pmatrix},\;\;
    \mathbf{D}^2\mathbf{s}=\begin{pmatrix} \mathbf{L}_0\mathbf{x}_0 \\ \mathbf{L}_1\mathbf{x}_1 \end{pmatrix}.
\end{equation}
Both $\mathbf{D}$ and $\mathbf{L}_{\mathcal{G}}$ are self-adjoint, so their eigenbases form complete orthonormal bases for spinors, inducing Fourier-like spectral representations. Let $\mathbf{U}$ ($\mathbf{U}_{\mathrm{H}}$) and $\mathbf{V}$ ($\mathbf{V}_{\mathrm{H}}$) collect the left and right singular vectors of $\mathbf{B}$ associated with its nonzero (zero) singular values. The Dirac operator and the super-Laplacian then admit the eigendecompositions $\mathbf{D}=\mathbf{\Phi}\mathbf{\Gamma}\mathbf{\Phi}^\top$ and $\mathbf{L}_{\mathcal{G}}=\mathbf{\Theta}\mathbf{\Lambda}\mathbf{\Theta}^\top$ \cite{bianconi2021topological}, with
\begin{equation}
\mathbf{\Phi} = \begin{pmatrix}
\mathbf{U} & \mathbf{0} & \mathbf{U}_{\mathrm{H}} & \mathbf{U} \\
-\mathbf{V} & \mathbf{V}_{\mathrm{H}} & \mathbf{0} & \mathbf{V}
\end{pmatrix}\!,\;\;
\mathbf{\Theta} = \begin{pmatrix}
\mathbf{U}_{\mathrm{H}} & \mathbf{0} & \mathbf{U} & \mathbf{0} \\
\mathbf{0} & \mathbf{V}_{\mathrm{H}} & \mathbf{0} & \mathbf{V}
\end{pmatrix}
\end{equation}
\begin{equation}
\mathbf{\Gamma} = \mathrm{blkdiag}\big(-\mathbf{\Sigma},\mathbf{0}_{\xi_1},\mathbf{0}_{\xi_0},\mathbf{\Sigma}\big)
\end{equation}
\begin{equation}
\mathbf{\Lambda} = \mathrm{blkdiag}\big(\mathbf{0}_{\xi_0},\mathbf{0}_{\xi_1},\mathbf{\Sigma}^2, \mathbf{\Sigma}^2\big)
\end{equation}
where $\mathbf{\Sigma}$ holds the nonnull singular values of $\mathbf{B}$, $\xi_0=\mathrm{dim}(\ker(\mathbf{B}))$, and $\xi_1=\mathrm{dim}(\ker(\mathbf{B}^\top))$.

\subsection{Dirac--Laplacian Frame Representation}

Comparing the two eigendecompositions, the Dirac operator best represents spinors whose node and edge components are strictly coupled along the same modes, while the Laplacian is preferable for domain-specific patterns. Since real signals generally mix \textit{spectrally coupled} and \textit{domain-specific} components, either basis alone is sub-optimal for compressed representation, motivating a redundant frame that concatenates both eigenbases together with a sparsity-enforcing representation.

\begin{proposition}[Dirac--Laplacian frame]
Let $\mathbf{\Phi}$ and $\mathbf{\Theta}$ denote the orthonormal eigenbases of the Dirac operator and of the super-Laplacian, respectively, and define $\mathbf{F} = (\mathbf{\Phi} \mid \mathbf{\Theta}) \in \mathbb{R}^{(V+E)\times 2(V+E)}$. Then $\mathbf{F}$ is a tight frame with frame bound $A=2$, i.e., $\mathbf{F}\mathbf{F}^\top = 2 \mathbf{I}_{V+E}$, and every spinor $\mathbf{s}\in\mathbb{R}^{V+E}$ satisfies the Parseval reconstruction formula $\mathbf{s} = \tfrac{1}{2} \sum_{i=1}^{2(V+E)} \langle \mathbf{f}_i , \mathbf{s} \rangle \mathbf{f}_i$, where $\mathbf{f}_i$ is the $i$-th column of $\mathbf{F}$.
\end{proposition}

While well-posed, this dictionary is limited to signals that are either perfectly compatible or perfectly incompatible between the two domains at a given scale. Handling signals entangled at different relative scales requires the more refined tools developed next.
\section{Learning Dirac Spectral Transforms}
\begin{figure*}
    \centering
    \includegraphics[width=1\linewidth]{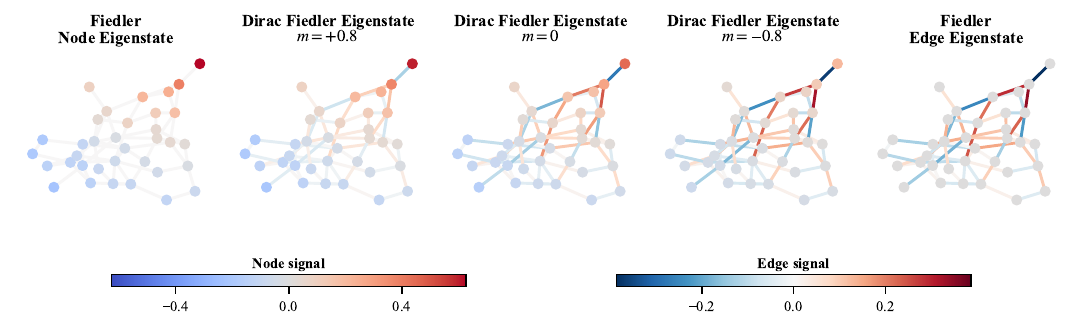}
    \caption{Effect of the mass parameter on Dirac eigenstates. The outer panels show the node (left) and edge (right) eigenstates at the smallest non-zero graph frequency. The central panel shows the Dirac eigenvector for the smallest positive eigenvalue.
    Intermediate panels trace the interpolation between coupled and uncoupled regimes as the mass varies.}
    \label{fig:diracEigenstates}
\end{figure*}
In this section, we first introduce a principled framework to build 
a spectral transform from mixtures of Dirac equations. We then leverage the proposed approach to formulate a learning problem that finds an optimal transform in a data-driven manner.
\subsection{Dirac equation of networks}

The Dirac equation on networks~\cite{bianconi2021topological} has been introduced for signal processing tasks to account for the relative scaling between topological modes, thereby enabling intermediate regimes between coupled and uncoupled dynamics \cite{wang2025dirac}. 
It builds on the eigenstates of the topological wave equation:
\begin{equation}\label{eq:DiracEquation}
\mathrm{i}\partial_t\mathbf{s}-\mathcal{H}\mathbf{s}=0,     
\end{equation}
where $\mathrm{i}$ denotes the imaginary unit, $\mathcal{H}=\mathbf{D}+m\begin{pmatrix}\mathbf{I}_V & \mathbf{0} \\\mathbf{0} & -\mathbf{I}_E \end{pmatrix}$ and $m$ parametrizes the \textit{mass} \cite{bianconi2024mass}.

The eigenstates $\boldsymbol{\psi}$ of $\mathcal{H}$, satisfying the equation 
$\mathcal{H}\boldsymbol{\psi} = \varrho\boldsymbol{\psi}$, 
can be shown to exhibit the following structure for each nonzero graph frequency $\lambda_i$\cite{wang2025dirac}:
\begin{equation}\label{eq:DiracEquationModes}
\psi_i^-=\zeta^-\begin{pmatrix}
        \frac{\lambda_i}{|\varrho_i|+m}\,\mathbf{u}_i \\
        -\mathbf{v}_i
    \end{pmatrix}
     ,\quad \psi_i^+=\zeta^+\begin{pmatrix}
        \mathbf{u}_i \\
        \frac{\lambda_i}{|\varrho_i|+m}\,\mathbf{v}_i
    \end{pmatrix} ,
\end{equation}
where $\zeta^{\pm}$ are constants used to have unit norm vectors.
The harmonic eigenvectors are invariants under this new parametrization and associated to energies equal in absolute value to the mass.
Fig.~\ref{fig:diracEigenstates} shows how the mass parameter allows to continuously interpolate between coupling and decoupling of the two domains, defining a more expressive way to interlace node and edge signals at different relative scales.

\subsection{Spectral Transform based on mixtures of Dirac equations}
Inspired by the structure of the eigenstates of the Dirac equation of networks in \eqref{eq:DiracEquationModes}, we propose a Fourier-like transform $\mathcal{T}_{\mathbf{D},\mathbf{m}}$  that incorporates a mode-dependent 
mass contribution across the natural modes of the Dirac operator. Specifically, let $\mathbf{m}=[m_1,\ldots,m_r]^\top\in \mathbb{R}^{r}$ be a vector of mass parameters, where $r = \frac{1}{2}[V + E - (\xi_0 + \xi_1)]$.
Following the same notation as in \eqref{eq:DiracEquationModes}, we define a mass-parameterized Fourier operator built as $\overline{\mathbf{\Psi}}(\mathbf{m})=({\mathbf{ \Psi}}^{-}(\mathbf{m})\mid\mathbf{\Psi}_\mathrm{H}\mid{\mathbf{\Psi}}^{+}(\mathbf{m}))$, where the $i$-th column $\overline{{\boldsymbol{\psi}}}_i$ of $\overline{\mathbf{\Psi}}(\mathbf{m})$ is built as
in \eqref{eq:DiracEquationModes}, but with a mode-specific mass parameter $m_i$.
Then, the proposed transform $\mathcal{T}_{\mathbf{D},\mathbf{m}}:\mathcal{C}^0\oplus\mathcal{C}^1\rightarrow\mathbb{R}^{V+E}$ and its inverse $\mathcal{T}_{\mathbf{D},\mathbf{m}}^{-1}:\mathbb{R}^{V+E}\rightarrow\mathcal{C}^0\oplus\mathcal{C}^1$ read as: 
\begin{align}
&\mathcal{T}_{\mathbf{D},\mathbf{m}}(\mathbf{s})=\overline{\mathbf{\Psi}}(\mathbf{m})^\top\mathbf{s}, \\
&\mathcal{T}_{\mathbf{D},\mathbf{m}}^{-1}(\hat{\mathbf{s}})=\overline{\mathbf{\Psi}}(\mathbf{m})\,\hat{\mathbf{s}}.
\end{align}
Intuitively, this transform is parametrized to encode, for each mode, the different degrees of coupling between the node and edge modes. 
As formulated, the mass-induced scaling symmetrically applies to chiral modes, i.e. the ones associated with eigenvalues with opposite sign, and applying the same mass parameter symmetrically within each chiral pair in \eqref{eq:DiracEquationModes} preserves orthogonality defining a complete orthonormal basis for spinors, which includes the Dirac and Laplacian eigenbases as the limiting regimes.

\subsection{Dirac-driven transform learning from signals}

In the sequel, we propose an effective algorithm to learn an orthogonal transform in a data-driven fashion. 
Building on the Dirac-based transform introduced above, we formulate a learning framework to jointly estimate the mode-specific mass parameters, which govern node–edge coupling, and enforce sparsity in the signal representation. We refer to the proposed framework as \textit{Dirac-driven transform learning} (DDTL).

We postulate the following signal model:
\begin{equation}\label{eq:SignalModel}
\mathbf{s}=\overline{\mathbf{\Psi}}(\mathbf{m}) \boldsymbol{\omega}
           +\mathbf{w},
\end{equation}
where $\boldsymbol{\omega}\in\mathbb{R}^{V+E}$ is the sparse vector of expansion coefficients and $\mathbf{w}$ is a zero-mean white Gaussian noise vector with covariance matrix $\sigma^2 \mathbf{I}_{V+E}$. Then, let us assume to collect $T$ topological spinors arranged in the data matrix 
$\mathbf{S} = [\mathbf{s}_1, \dots, \mathbf{s}_T]$, and denote by 
$\mathbf{\Omega} = [\boldsymbol{\omega}_1, \dots, \boldsymbol{\omega}_T]$ 
their corresponding coefficient vectors in the parametrized basis to be learned. 
Under the model in \eqref{eq:SignalModel}, the maximum likelihood 
estimator yields the following least-squares problem:
\begin{equation}
    \underset{\mathbf{m}, \mathbf{\Omega}}{\mathrm{min}}
    \quad
    \|\mathbf{S} - \overline{\mathbf{\Psi}}(\mathbf{m}) \mathbf{\Omega}\|_{\mathrm{F}}^2.
\end{equation}
\begin{figure*}[t!]
    \centering    \includegraphics[width=0.99\linewidth]{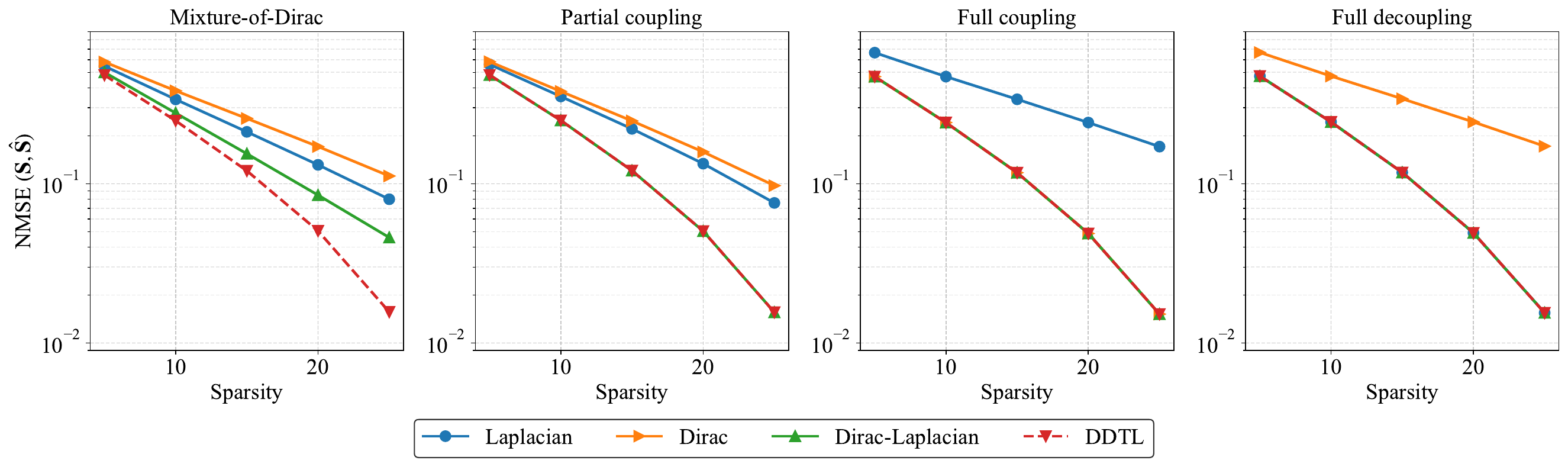}
    \caption{Synthetic results for sparsity-reconstruction trade-off with different classes of signals on different bases \vspace{-.2cm}}
    \label{fig:fig_0}
    \label{fig:placeholder}
\end{figure*}
Since the mass parameters affect the scaling of the eigenmodes in a nonlinear 
manner, we adopt a reparametrization strategy: instead of learning the mass parameter $m_i$ directly, we learn its induced scaling factor, defined as
$k_i = \frac{\lambda_i}{|\varrho_i| + m_i}$, for all $i=1,\ldots,r$, which enters the model linearly. Under this reparametrization, the resulting 
objective function becomes biconvex in the variables $\mathbf{k}=[k_1,\ldots,k_r]^T$ and $\mathbf{\Omega}$, being linear with respect to each of the two variables holding the other fixed.
To promote group sparsity in the representation coefficients, we restrict 
the feasible set of codes to
$\mathcal{B}_{\eta_0}
:=
\left\{
\mathbf{\Omega} \;:\; \|\mathbf{\Omega}\|_{2,0} = \eta_0
\right\},$
i.e., the set of matrices with exactly $\eta_0$ nonzero rows. 
Furthermore, to prevent scale ambiguities related to the nonlinear effect of the mass that may interfere with the 
localization properties of the learned modes, we constrain the columns 
of $\overline{\mathbf{\Psi}}(\mathbf{m})$ to be normalized. This is equivalent to imposing $\overline{\mathbf{\Psi}}(\mathbf{m}) \in \mathrm{Ob}(V+E, V+E)$,
where $\mathrm{Ob}(V+E, V+E)$ denotes the oblique manifold, i.e., the set of matrices with unit-norm columns. 
We transfer these structural constraints to auxiliary splitting variables $\mathbf{P},\mathbf{X}$, ensuring solutions with the desired structures decoupling them from the block-wise convexity of the learning problem.  
By additionally defining a feasible set for $\mathbf{k}$, we finally obtain 
the following optimization problem, where $\eta_0 \in \mathbb{N}$ and $c_1,c_2 \in [0,1]$ act as tunable hyperparameters:
\begin{equation}\label{eq:LearningProblem}
\begin{aligned}
\underset{\mathbf{k},\mathbf{\Omega},\mathbf{P},\mathbf{X}}{\mathrm{min}} \quad & \|\mathbf{S}-\overline{\mathbf{\Psi}}(\mathbf{k})\mathbf{\Omega}\|_\F^2 \nonumber \\
    \mathrm{s.t.} \quad & -c_2\mathbf{1} \leq \mathbf{k}\leq c_1\mathbf{1} \nonumber \\
    & \mathbf{X} \in\mathcal{B}_{\eta_0} \nonumber \\
    & \mathbf{P} \in \mathrm{Ob}(V+E,V+E) \nonumber \\
    & \mathbf{P}-\overline{\mathbf{\Psi}}(\mathbf{k})=\mathbf{0}_{} \nonumber \\
    & \mathbf{X}-\mathbf{\Omega}=\mathbf{0} \nonumber
\end{aligned}
\tag{P1}
\end{equation}
To solve \eqref{eq:LearningProblem}, we resort to the alternating direction method of multipliers (ADMM) \cite{boyd2011distributed}, based on the constrained minimization of the augmented Lagrangian. To this end, we introduce two matrix Lagrange multipliers $\mathbf{H}$ and $\mathbf{M}$, yielding
\begin{equation}
\begin{aligned}
\mathcal{L} &= \|\mathbf{S} - \overline{\mathbf{\Psi}}(\mathbf{k})\mathbf{\Omega}\|_{\mathrm{F}}^2 + \frac{\rho_1}{2}
\|\overline{\mathbf{\Psi}}(\mathbf{k}) - \mathbf{P} + \mathbf{H}\|_{\mathrm{F}}^2 \\
&\quad + \frac{\rho_2}{2}
\|\mathbf{\Omega} - \mathbf{X} + \mathbf{M}\|_{\mathrm{F}}^2.
\end{aligned}
\end{equation}
Combining the feasibility constraints defined in \eqref{eq:LearningProblem} with the augmented Lagrangian formulation above leads to the following iterative updates\footnote{For notational simplicity, we omit time indices, with the understanding that the scheme proceeds sequentially.}: 
\begin{equation}
\begin{aligned}
\mathbf{k}^{t+1}=\;&\underset{-c_2\mathbf{1} \leq \mathbf{k}\leq c_1\mathbf{1}}{\mathrm{argmin}}
\; \|\mathbf{S}-\overline{\mathbf{\Psi}}(\mathbf{k})\mathbf{\Omega}\|_\F^2 + \frac{\rho_1}{2}
\|\overline{\mathbf{\Psi}}(\mathbf{k}) - \mathbf{P} + \mathbf{H}\|_{\mathrm{F}}^2 \\
\mathbf{\Omega}^{t+1}=\;& \Big[2\overline{\mathbf{\Psi}}(\mathbf{k})^\top\overline{\mathbf{\Psi}}(\mathbf{k}) + \rho_2\mathbf{I}\Big]^{-1}
\big[\rho_2(\mathbf{X}-\mathbf{M}) + \overline{\mathbf{\Psi}}^\top(\mathbf{k})\mathbf{S}\big] \\
\mathbf{P}^{t+1}=\;& \mathcal{R}_{\mathrm{Ob}}\Big\{\mathbf{H}+\overline{\mathbf{\Psi}}(\mathbf{k})\Big\} \\
\mathbf{X}^{t+1}=\;& \mathcal{R}_{\mathcal{B}_{\eta_0}}\Big\{\mathbf{\Omega}+\mathbf{M}\Big\} \\
\mathbf{H}^{t+1}=\;& \mathbf{H} + [\overline{\mathbf{\Psi}}(\mathbf{k}) - \mathbf{P}] \\
\mathbf{M}^{t+1}=\;& \mathbf{M} +(\mathbf{\Omega}-\mathbf{X}) \nonumber
\end{aligned}
\end{equation}
The retractions $\mathcal{R}_{\mathrm{Ob}}$ and 
$\mathcal{R}_{\mathcal{B}_{\eta_0}}$ are implemented, respectively, 
by normalizing the columns of the argument matrix 
\cite{absil2008optimization}, and by retaining only the $\eta_0$ rows 
with largest $\ell_2$ norm (i.e., highest energy) while setting all 
remaining rows to zero \cite{baraniuk2010model}. 
The update step with respect to $\mathbf{k}$ instead amounts to a  convex optimization problem, which can be solved efficiently using numerical methods for quadratic programming.
The convergence of the algorithm is monitored using standard arguments from \cite{boyd2011distributed}. 
In practice, the algorithm performs well both with random initialization and in limit-regime initializations as the Dirac and Laplacian eigenbases are.

\section{Numerical results}

In the sequel, we assess the performance of the proposed method via numerical experiments on synthetic and real data.

\begin{figure}[t]
    \centering
    \includegraphics[width=0.95\linewidth]{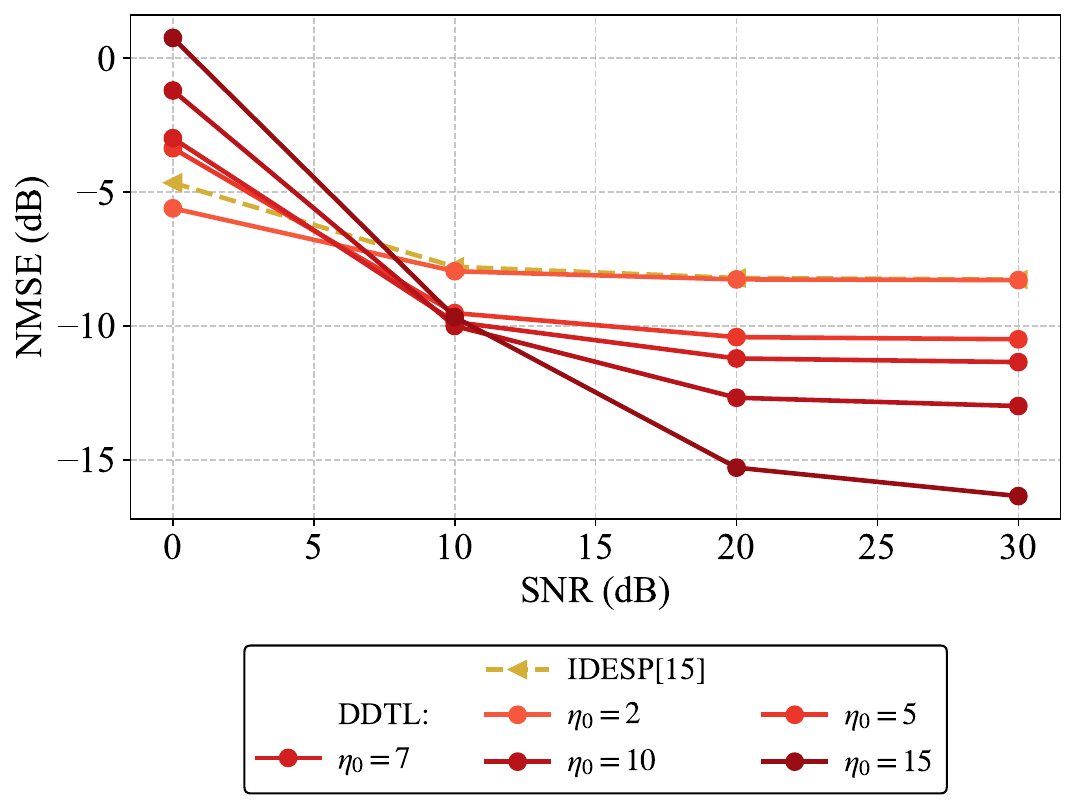}
    \caption{NMSE vs. SNR, for different algorithms.}
    \label{fig:fig_1}
\end{figure}

\noindent\textbf{Sparse Representation.} With this experiment, we validate the considered representation basis for different classes of signals. We considered instances of Erdős–Rényi random graphs with $V=40$ nodes and $E=80$ edges. We consider the following classes of signals all derived from the considered model \eqref{eq:SignalModel}: \emph{(i) fully coupled signals} (Dirac regime), obtained for $m_i=0, \forall i\in[r]$, \emph{(ii) fully decoupled signals} (Laplacian regime), obtained for $m_i>>0, \forall i \in[r]$, \emph{(iii) partially coupled signals} (Dirac-Laplacian regime), obtained for $m_i = 0, \forall i\in\mathcal{I}_{\mathcal{C}},m_j>>0, \forall j \in\mathcal{I}_{\mathcal{U}}$ such that $\mathcal{I}_{\mathcal{C}}\cup\mathcal{I}_{\mathcal{U}}=[r]$, \emph{(iv) mixture-of-Dirac signals} where in \eqref{eq:SignalModel} we define a mass distribution across the modes via a radial basis function with a Cauchy-like kernel. 
Ideally, coupling should degrade with frequency as high frequency patterns are associated with localized modes.
We fix the bandwidth to $\eta_0 = 35$ and assume a shared spectral support during the generation step. In each scenario, we generate $M = 600$ signals with random Gaussian coefficients. Harmonic components are excluded, as they are optimally captured by any of the considered bases. We solve the DDTL problem on the given signals setting $c_1=c_2=1, \rho_1=\rho_2=10$, and then compare the normalized mean squared error (NMSE) between the observed data $\mathbf{S}$ and their reconstructed version $\hat{\mathbf{S}}$, obtained via orthogonal matching pursuit (OMP) \cite{tropp2007signal}, across different sparsity levels. The comparison is performed over the considered bases: the Laplacian basis, the Dirac basis, the joint Dirac–Laplacian frame, and the DDTL transform.\\
In Figure~\ref{fig:fig_0}, we show the results averaged over 10 realizations of the depicted experimental setting. As expected, we can observe a factor-of-two in bandwidth relation between Dirac and Laplacian representation in the fully coupled and decoupled signals, with the frame achieving optimal sparsity-reconstruction trade-off in the two scenarios and in the partially coupled one. The proposed data-driven transform not only achieves optimal representation performance in these three scenarios, but also exhibits a significant gain in more challenging settings (i.e., mixture-of-Dirac signals), effectively capturing complex cross-domain entanglement patterns that cannot be described by purely Dirac or Laplacian regimes.\\
\noindent\textbf{Denoising.} In this experiment, we evaluate the performance of our proposed method in a denoising task, to assess the impact of noise on the learning algorithm. 
We consider topological signals naturally entangled as quantities measured in the context of water distribution networks (WDNs), where node and edge signals represent, respectively, pressure and flow data.
This induces a challenging model mismatch as the signals cannot be assumed to be modeled with a perfectly finite bandwidth.
We consider the Anytown network, consisting of $V=22$ nodes and $E=41$ edges, retrieving the first $T=240$ node pressure measurements and pipe flow-rates as collected in \cite{tello2024large}.
We thus have a ground truth data matrix $\mathbf{S}$ that we corrupt with AWGN, at various signal-to-noise ratio (SNR) levels.
The proposed DDTL problem is solved for different bandwidths, considering the solution as the optimal filtered signal to be $\overline{\mathbf{\Psi}}(\mathbf{k}^*)\mathbf{\Omega}^*\approx\mathbf{S}$.
We compare the performance of the algorithm with the iterative filtering framework (IDESP) proposed in \cite{wang2025dirac}. The results are reported in Fig.~\ref{fig:fig_1}, which depicts the NMSE between the ground-truth and filtered signals as a function of SNR, averaged over 10 independent noise realizations.
The bandwidth plays a crucial role in noise robustness. For small bandwidth values, the proposed method slightly outperforms IDESP at low SNR levels, whereas increasing the bandwidth as the SNR improves yields progressively larger performance gains. These results demonstrate the capability of the proposed model to learn an informative basis for representing structured signals even under mild modeling assumptions. On the other hand, the algorithm relies on data-driven learning, whereas IDESP constitutes a fully analytical alternative.

\section{Conclusions}
In this paper, we proposed a learnable spectral transform for the representation of topological signals jointly defined on nodes and edges to overcome the limitations of fixed representations.
We introduced a parametric Dirac-based transform with mode-dependent mass parameters, enabling a continuous interpolation between coupled and decoupled regimes. We then formulated a Dirac-Driven Transform Learning (DDTL) framework to learn these parameters from data under sparsity constraints. Numerical results on synthetic and real-world datasets demonstrated improved sparsity–distortion trade-offs and competitive denoising performance compared to Laplacian-, Dirac-, and frame-based representations. Future work will investigate extensions to higher-order structures and integration within learning-based architectures for topological signal processing.

\begingroup
\balance
\bibliographystyle{bibstile}
\bibliography{bibliography}

@article{bianconi2021topological,
  title={The topological {D}irac equation of networks and simplicial complexes},
  author={Bianconi, G.},
  journal={Journal of Physics: Complexity},
  volume={2},
  number={3},
  year={2021},
  publisher={IOP Publishing}
}

@article{wang2025dirac,
  title={Dirac-equation signal processing: {P}hysics boosts topological machine learning},
  author={Wang, R. and Tian, Y. and Li{\`o}, P. and Bianconi, G.},
  journal={PNAS nexus},
  volume={4},
  number={5},
  pages={139},
  year={2025},
  publisher={Oxford University Press US}
}

@article{battiloro2024generalized,
  title={Generalized simplicial attention neural networks},
  author={Battiloro, C. and Testa, L. and Giusti, L. and Sardellitti, S. and Di Lorenzo, P. and Barbarossa, S.},
  journal={IEEE Transactions on Signal and Information Processing over Networks},
  year={2024},
  publisher={IEEE}
}

@book{grady2010discrete,
  title={Discrete calculus: {A}pplied analysis on graphs for computational science},
  author={Grady, L.J. and Polimeni, J.R.},
  volume={3},
  year={2010},
  publisher={Springer}
}

@article{sandryhaila2013discrete,
  title={Discrete {S}ignal {P}rocessing on {G}raphs},
  author={Sandryhaila, A. and Moura, J.M.F.},
  journal={IEEE {T}ransactions on {S}ignal {P}rocessing},
  volume={61},
  number={7},
  pages={1644--1656},
  year={2013},
  publisher={IEEE}
}

@inproceedings{smola2003kernels,
  title={Kernels and {R}egularization on {G}raphs},
  author={Smola, A.J. and Kondor, R.},
  booktitle={16th Annual Conference on Learning Theory and 7th Kernel Workshop, COLT/Kernel 2003, Washington, DC, USA},
  pages={144--158},
  year={2003},
  organization={Springer}
}

@article{Barbarossa_2020,
   title={Topological Signal Processing Over Simplicial Complexes},
   volume={68},
   ISSN={1941-0476},
   url={http://dx.doi.org/10.1109/TSP.2020.2981920},
   DOI={10.1109/tsp.2020.2981920},
   journal={IEEE Transactions on Signal Processing},
   publisher={Institute of Electrical and Electronics Engineers (IEEE)},
   author={Barbarossa, S. and Sardellitti, S.},
   year={2020},
   pages={2992–3007} }

@article{isufi2025topological,
  title={Topological signal processing and learning: Recent advances and future challenges},
  author={Isufi, E. and Leus, G. and Beferull-Lozano, B. and Barbarossa, S. and Di Lorenzo, P.},
  journal={Signal Processing},
  pages={109930},
  year={2025},
  publisher={Elsevier}
}

@article{carletti2025global,
  title={Global topological {D}irac synchronization},
  author={Carletti, T. and Giambagli, L. and Muolo, R. and Bianconi, G.},
  journal={Journal of Physics: Complexity},
  volume={6},
  number={2},
  pages={025009},
  year={2025},
  publisher={IOP Publishing}
}

@article{calmon2023local,
  title={Local {D}irac synchronization on networks},
  author={Calmon, L. and Krishnagopal, S. and Bianconi, G.},
  journal={Chaos: An Interdisciplinary Journal of Nonlinear Science},
  volume={33},
  number={3},
  year={2023},
  publisher={AIP Publishing}
}

@inproceedings{yang2022simplicial,
  title={Simplicial convolutional neural networks},
  author={Yang, M. and Isufi, E. and Leus, G.},
  booktitle={IEEE International Conference on Acoustics, Speech and Signal Processing},
  pages={8847--8851},
  year={2022},
  organization={IEEE}
}

@inproceedings{mey2024dirac,
  title={Dirac--{B}ianconi Graph Neural Networks-Enabling long-range graph predictions},
    author={Nauck, C. and Gorantla, R. and Lindner, M. and Schürholt, K. and Mey, A.S.J.S. and Hellmann, F.},
  booktitle={ICML 2024 {W}orkshop on {G}eometry-grounded {R}epresentation {L}earning and {G}enerative {M}odeling},
  year={2024}
}

@article{calmon2023dirac,
  title={Dirac signal processing of higher-order topological signals},
  author={Calmon, L. and Schaub, M.T. and Bianconi, G.},
  journal={New Journal of Physics},
  volume={25},
  number={9},
  pages={093013},
  year={2023},
  publisher={IOP Publishing}
}

@article{wee2023persistent,
  title={Persistent {D}irac for molecular representation},
  author={Wee, J. and Bianconi, G. and Xia, K.},
  journal={Scientific Reports},
  volume={13},
  number={1},
  pages={11183},
  year={2023},
  publisher={Nature Publishing Group UK London}
}

@book{absil2008optimization,
  title={Optimization algorithms on matrix manifolds},
  author={Absil, P-A. and Mahony, R. and Sepulchre, R.},
  year={2008},
  publisher={Princeton University Press}
}

@book{chung1997spectral,
  title={Spectral graph theory},
  author={Chung, F.R.K.},
  volume={92},
  year={1997},
  publisher={American Mathematical Soc.}
}

@article{boyd2011distributed,
  title={Distributed optimization and statistical learning via the alternating direction method of multipliers},
  author={Boyd, S. and Parikh, N. and Chu, E. and Peleato, B. and Eckstein, J. and others},
  journal={Foundations and Trends{\textregistered} in Machine learning},
  volume={3},
  number={1},
  pages={1--122},
  year={2011},
  publisher={Now Publishers, Inc.}
}

@article{baraniuk2010model,
  title={Model-based compressive sensing},
  author={Baraniuk, R.G. and Cevher, V. and Duarte, M.F. and Hegde, C.},
  journal={IEEE Transactions on {I}nformation {T}heory},
  volume={56},
  number={4},
  pages={1982--2001},
  year={2010},
  publisher={IEEE}
}

@book{ortega2022introduction,
  title={Introduction to graph signal processing},
  author={Ortega, A.},
  year={2022},
  publisher={Cambridge University Press}
}

@book{bianconi2021higher,
  title={Higher-order networks},
  author={Bianconi, G.},
  year={2021},
  publisher={Cambridge University Press}
}

@article{baccini2022weighted,
  title={Weighted simplicial complexes and their representation power of higher-order network data and topology},
  author={Baccini, F. and Geraci, F. and Bianconi, G.},
  journal={PHYSICAL REVIEW E Phys Rev E},
  volume={106},
  pages={034319},
  year={2022},
  publisher={American Physical Society}
}

@article{tropp2007signal,
  title={Signal recovery from random measurements via orthogonal matching pursuit},
  author={Tropp, J.A. and Gilbert, A.C.},
  journal={IEEE Transactions on information theory},
  volume={53},
  number={12},
  pages={4655--4666},
  year={2007},
  publisher={IEEE}
}

@article{bianconi2024mass,
  title={The mass of simple and higher-order networks},
  author={Bianconi, G.},
  journal={Journal of Physics A: Mathematical and Theoretical},
  volume={57},
  number={1},
  year={2024},
  publisher={IOP Publishing}
}

@article{tello2024large,
  title={Large-scale multipurpose benchmark datasets for assessing data-driven deep learning approaches for water distribution networks},
  author={Tello, A. and Truong, H. and Lazovik, A. and Degeler, V.},
  journal={Engineering Proceedings},
  volume={69},
  number={1},
  pages={50},
  year={2024},
  publisher={MDPI}
}
\endgroup
\end{document}